# Yeast caspase 1 suppresses the burst of reactive oxygen species and maintains mitochondrial stability in *Saccharomyces cerevisiae*

*[1]Lin DU, [1]Xiaodan HUANG, [1]Jian TAN, [2]Yongjun LU, [2]Shining ZHOU,

(1 Food Science College, Zhongkai University of Agriculture and Engineering, Guangzhou 510225, China, 2 Department of Biochemistry, School of Life Sciences, Sun Yat-sen University, Guangzhou 510275, China)
Email: dulin2000@163.com

Abstract: Caspases are a family of cysteine proteases that play essential roles during apoptosis, and we presume some of them may also protect the cell from oxidative stress. We found that the absence of yeast caspase 1（Yca1）in *Saccharomyces cerevisiae* leads to a more intense burst of mitochondrial reactive oxygen species (ROS). In addition, compared to wild type yeast cells, the ability of *yca1* mutant cells to maintain mitochondrial activity is significantly reduced after either oxidative stress treatment or aging. During mitochondrial ROS burst, deletion of the *yca1* gene delayed structural damage of a green fluorescent protein (GFP) reporter bound in the inner mitochondrial membrane. This work implies that yeast caspase 1 is closely connected to the oxidative stress response. We speculate that Yca1 can discriminate proteins damaged by oxidation and accelerate their hydrolysis to attenuate the ROS burst.

## 1. Introduction

The unicellular yeast *Saccharomyces cerevisiae* is a useful model to elucidate the molecular mechanisms underlying programmed cell death (PCD) pathways. *S. cerevisiae* PCD shares many morphological and biochemical features with mammalian apoptosis, including chromatin condensation and nuclear DNA fragmentation. One hallmark of apoptosis in mammalian cells is the induction of caspases. Caspases are direct regulators of both cytokine-induced and stress-induced cell apoptosis that act as proteases that initiate and execute cell death through the degradation of cellular components (Nicholson 1999; Alenzi, Lotfy et al. 2010). Yeast contains only one gene homolog of caspases, named *YCA1*, encoding the metacaspase yeast caspase 1 (Madeo, Herker et al. 2002). This metacaspase adopts a caspase-like fold, with active site loops arranged similarly to other caspases (Wong, Yan et al. 2012), but with different substrate specificity than caspases (Wilkinson and Ramsdale 2011). Glyceraldehyde-3-phosphate dehydrogenase was the first identified Yca1-specific substrate degraded during $H_2O_2$-induced programmed cell death (Silva, Almeida et al. 2011). Recent work has indicated that some caspases may play roles in other processes in addition to their death functions (Yi and Yuan 2009; Shrestha and Megeney 2012). Many genes involved in stress response, instead of genes encoding effectors or markers of apoptosis, exhibit a particular transcriptional response prior to the initiation of apoptosis (Munoz, Wanichthanarak et al. 2012). The roles of Yca1 in the mechanism of yeast PCD and in other stress response processes remain poorly understood.

Both in yeast and in mammalian PCD, mitochondria play a major role in the final determination of cell fate. The extent to which yeast PCD resembles and precedes apoptotic death in multicellular organisms or is a distinct form of PCD remains unknown. Some researchers have proposed that the programmed cell death in unicellular yeast cells may serve to eliminate unwanted cells or to increase the health and adaptation ability of the population (Herker et al., 2004), though the evolutionary benefit of a cellular suicide program in unicellular organism remains poorly understood.

In yeast, reactive oxygen species (ROS) production usually precedes caspase activation (Madeo, Herker et al. 2004; Perrone, Tan et al. 2008). In our previous work, we found that a concentration of 30 mM formic acid results in a rapid burst of intracellular ROS in mitochondria. Interestingly, formic acid treatment of logarithmic phase cells resulted in a higher level of ROS in *yca1* mutant cells than that produced in wild type cells, suggesting a role for Yca1 in anti-oxidation (Du, Su et al. 2008).

We propose that in *S.cerevisiae*, Yca1 may affect mitochondrial activity and attenuate the generation or diffusion of ROS. In this work, we inhibit the activity of Yca1 and then observe the effects of this inhibition on the burst of ROS. We hypothesize that the aspartic enzyme activity of Yca1 promotes the fast degradation of oxidized proteins during ROS burst (Du, Lu et al. 2009; Du 2010).

**2. Materials and Methods**

2.1 Yeast strains, plasmids and medium

We used the *S. cerevisiae* strain BY4741 and a derivative of BY4741 in which the *YCA1* gene was deleted. This strain contains no pigment to interfere in fluorescence assays. Wild-type *Saccharomyces cerevisiae* BY4741 and the *yca1* mutant strain were kind gifts from Professor David Goldfarb of University of Rochester. Cells were grown in YPD medium (1% yeast extract, 2% Bacto peptone and 2% glucose), or in selective synthetic and complete media.

2.2 Fusion of GFP with mitochondrial protein

A pYES-mtGFP plasmid, provided by Westermann and Neupert (Westermann and Neupert 2002), expressed a galactose-inducible, mitochondrial ATPase-targeted green fluorescent protein (GFP). This plasmid was transformed into yeast BY4741 wild type and *yca1* mutant cells using the LiAc protocol (Guthrie and Fink, 1991).

2.3 Observation of intracellular ROS with confocal microscopy

The cells were incubated with 5 μg/mL dihydrorhodamine 123 (DHR123, Sigma-Aldrich) or 20 μM 2', 7'- dichlorofluorescin diacetate (DCFH-DA, Cayman Co.) for 10 min at 37°C. After formic acid treatment, cells were placed on a slide. Confocal microscopy (Leica TCS-SP2) was used to observe the cells using an excitation wavelength of 488nm and an emission wavelength of 525nm.

2.4 Flow cytometry analysis of ROS level during mitochondrial ROS burst

Wild type BY4741 and the *yca1* mutant were cultured for 6h (30°C, 200 r / min shaking) to early logarithmic phase. 2μL of 10 mM DCFH-DA dissolved in DMSO was added to 1 mL cell culture, for a final concentration of 20μM DCFH-DA. After incubation in the dark for 5 min, samples were treated with 15 μM of broad-spectrum caspase inhibitor Z-VAD-FMK（from Promega）and incubated in the dark for 10 min. The same volume of DMSO was added to the control sample. Formic acid was quickly added to each sample, and cells were again incubated in the dark. At different time points after formic acid addition, 100μL of each sample was combined with 900μL chilled PBS, transferred to FACS tubes, and analyzed immediately by flow cytometry. After analysis of all samples, re-analysis of samples was carried out in the same sequence. The excitation wavelength of FACS was 488 nm and the emission wavelength was 530 nm.

In order to detect superoxide anion, we used 10 mM of the fluorescence probe Dihydroethidium (DHE, dissolved in DMSO). 0.5μL of this probe solution was added to 1 mL cell culture; the final concentration of DHE was 5μM. Cells were incubated in the dark for 15 min at 30 °C, treated with different concentrations of formic acid in the dark, and then analyzed by flow cytometry. The excitation wavelength was 535 nm, and the emission wavelength was 610 nm.

2.5 Detection of mitochondrial membrane potential

1μL of 1mM Rhodamine 123 (RH123) was added to 1mL yeast cell suspension and observed with confocal microscopy or analyzed with flow cytometry immediately to detect the potential change of the inner mitochondrial membrane ($\Delta\Psi m$). The excitation wavelength was 488 nm and the emission wavelength was 530 nm.

3. Results

***YCA1* gene deletion does not affect the potential of mitochondrial membrane or the mitochondrial production of superoxide anion in logarithmic phase cells**

Formic acid has been shown to inhibit the mitochondrial cytochrome c oxidase (Liesivuori and Savolainen 1991). Our previous work showed that formic acid can induce a rapid ROS burst in logarithmically-growing yeast cells, and that mitochondria are the major producer of ROS (Du, Su et al. 2008). Here, we used formic acid to induce intracellular ROS burst and determine the effect of Yca1 on the process of ROS burst in cells.

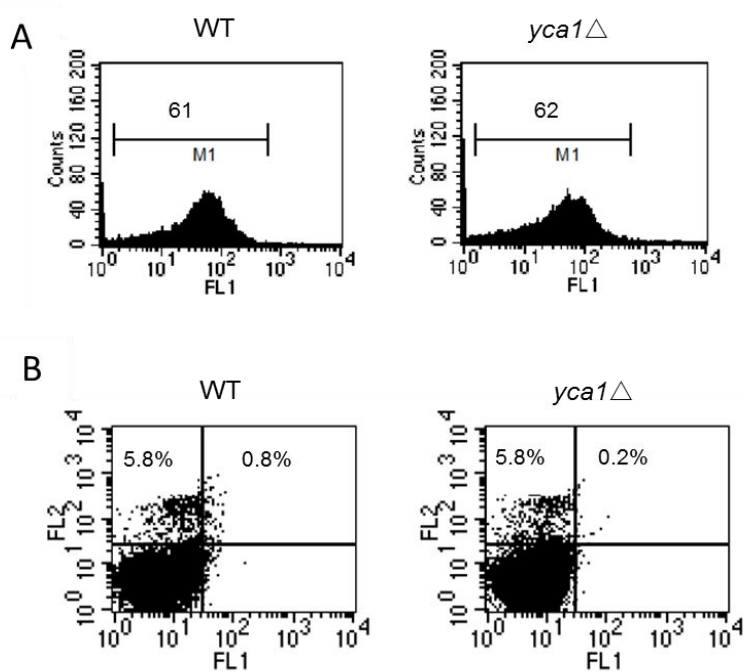

Fig. 1 A. Mitochondrial membrane potential of wild type and *yca1*Δ S.cerevisiae cells in log growth phase. The cells were stained by RH123 and analyzed with FACS.

B. The generation of superoxide anion induced by formic acid in wild type and *yca1*Δ cells of *S.cerevisiae*. The cells were stained by DHE and analyzed with FL2 channel of FACS.

Because the generation of ROS is affected by mitochondrial activity, mitochondrial activity of wild type and *YCA1* mutant strains were first measured before induction of ROS burst. The production of ROS by mitochondria results in structural damage to the membrane, leading to a decline in mitochondrial membrane potential (Zorov, Juhaszova et al. 2006). Thus, mitochondrial membrane potential reflects the activity of mitochondria and is an indicator of ROS generation. In order to remove the effect of mitochondria aging and cell aging, wild type and *yca1* mutant cells were cultured to logarithmic phase (6 hour). Next, the cells were labeled with fluorescent dye RH123 to compare mitochondrial activities quantitatively. We found that in logarithmic phase, the mitochondrial membrane potentials of wild type and *yca1* mutant cells are almost the same (Fig 1A).

One form of ROS generated from the mitochondrial electron transport chain is superoxide anion (Simon, Haj-Yehia et al. 2000). In order to compare superoxide anion production in wild type and *yca1* mutant cells after formic acid treatment, the superoxide anion fluorescent dye DHE was used. As shown in Fig 1B, there is no significant difference between the superoxide anion positive percentages of wild type and mutant cells (6.6% and 6.0%), suggesting that both type of cells have similar ability to generate superoxide anion after formic acid treatment.

Therefore, whether measured by mitochondrial membrane potential or by the ability to produce superoxide anion with formic acid treatment, the mitochondrial activity of log phase *yca1* mutant cells does not differ from that of wild type cells.

**Inhibition of Yca1 activity or knockout of *YCA1* gene intensifies the ROS burst in cytoplasm**

DHR123, which can be oxidized into fluorescent RH123, is widely used to detect ROS. After labeling with DHR123 and treatment of 50mM formic acid for 40 min, both wild type and *yca1* mutant cells showed intense fluorescence, indicating a high level of ROS (Fig 2A). The average fluorescence in wild type cells seemed stronger than that in the *yca1* mutant cells. However, the intensity of DHR123 fluorescence is determined by two factors: the level of ROS and the membrane potential of mitochondria. Therefore, a stronger fluorescent signal from DHR123 in the wild type cells may not indicate a higher level of ROS. We speculate that this stronger fluorescence is more likely due to the lower level of ROS, less mitochondrial damage, and higher mitochondrial membrane potential in the wild type cells, such that the higher mitochondrial membrane potential results in stronger fluorescence of DRH123.

Fig. 2 A. ROS production induced by formic acid in wild type and *yca1Δ* cells of *S.cerevisiae*. The cells were treated with 50 mM formic acid and stained by 5μg/mL DHR123 and then analyzed by confocal microscopy (Bar, 5μm) and FACS.

B. ROS level in cytoplasm with formic acid treatment in wild type and *yca1Δ* cells. The cells were stained by DCFH-DA in PBS and treated with 50mM formic acid for approximately 40 min and then analyzed with confocal microscopy (Bar, 10μm) and FACS. The experiment was performed at least three times, with similar results.

C. ROS burst process induced by 50mM formic acid in wild type and *yca1Δ* cells stained by DCFH-DA and incubated in medium. Samples were analyzed by FACS. The experiment was performed at least three times, with similar results.

Unlike DHR123, DCFH-DA is mainly distributed in the cytoplasm, so its fluorescence intensity is a better indicator of ROS level in cells. *yca1* mutant cells, labeled by DCFH-DA and treated by 50mM formic acid for about 20 min, showed stronger fluorescence of ROS than wild-type cells, indicating that Yca1 seemed to affect the burst of ROS (Fig 2B).

Intracellular ROS levels change quickly during ROS burst, so we analyzed the process of ROS burst in a time-dependent manner for wild type and *yca1* null cells. The broad-spectrum caspase inhibitor Z-VAD-FMK was used to inhibit the activity of Yca1. After 50mM formic acid treatment, each sample showed rapid ROS production. At the beginning of detection (5 min), the percentage of ROS positive cells in wild-type cells sample is higher than those in the other three samples (*yca1* mutant cells, wild type cells plus Z-VAD-FMK, and *YCA1* gene mutant cells plus Z-VAD-FMK). However, the proportion of ROS positive wild type cells did not increase significantly with time, while in the other three samples the percentage of ROS positive cells increased much faster. After 20 min, most cells in these samples were ROS positive but the low level of positive cells in the wild type sample persisted. Thus, the knockout of yeast caspase 1 gene, or the inhibition of Yca1 enzyme activity, increased the ROS level in cytoplasm during ROS burst (Fig 2C). We compared the death rates of the two strains and found that the death rate of *yca1* mutant cells was also higher than that of wild type cells (data not shown).

**YCA1 gene deletion results in reduced mitochondrial activity during ROS burst and cell aging**

There are some similarities between ROS burst and the cell aging process, both of which cause oxidative damage. We hypothesized that in aged cells, the deletion of *YCA1* gene would result in a faster reduction of mitochondrial activity. Thus, we compared the mitochondrial activities of aged wild type cells and *yca1* mutant cells.

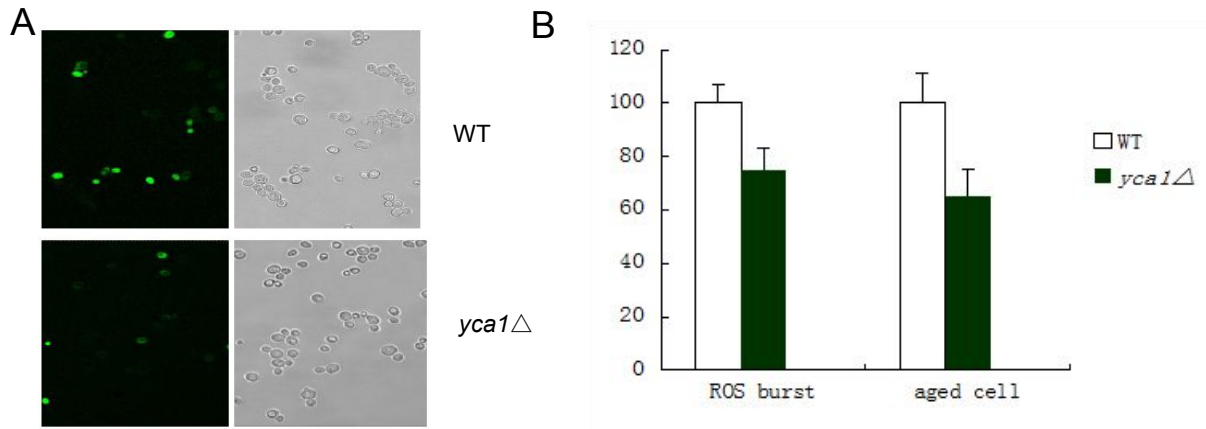

Fig. 3 A. Comparison of mitochondrial membrane potential between aged wild type and *yca1*Δ cells with confocal microscope observation. These aged cells were cultured for 72 hours, and the mitochondrial membrane potential was measured with RH123 staining (Bar, 5μm.)

B. The activity of mitochondrial membrane potential of wild type and *yca1*Δ cells after ROS burst or cell aging. The wild type and *yca1*Δ cells were treated with 60 mM formic acid for 40 min to induce ROS burst. After cells were washed with PBS, mitochondrial membrane potential was measured with RH123 staining and FACS analysis. The aging cells cultured for 72 hour were analyzed with the same method.

Mitochondrial membrane potentials of wild type and *yca1* mutant cell were detected after formic acid treatment and RH123 staining. The experiment result shows that, after 72h culture, the fluorescence of RH123, an indicator of mitochondrial membrane potential and mitochondrial activity, is significantly higher in wild type cells (Fig 3A). Quantitative analysis also showed that the decline of mitochondrial activity in *yca1* mutant cells was faster in the aging process (Fig 3B). Thus, *YCA1* deletion decreases the ability of the cell to maintain mitochondrial activity.

Based on the above finding that ROS levels in the wild type cells are relatively lower during ROS burst, we conclude that a lower level ROS results in less mitochondrial damage. Therefore the mitochondrial activity of wild type cell is relatively higher than that of the *yca1* mutant. The result that *yca1* mutant cells possess less mitochondrial activity than wild type cells after logarithmic phase may explain the finding that *yca1* mutant cells in stationary phase are more resistant to hydrogen peroxide than wild type cells (Madeo, Herker et al. 2002; Pereira, Silva et al. 2008). Additionally, we found that during stationary phase, ROS levels in *yca1* mutant cells were lower than in the wild type control after the same formic acid treatment (data not shown).

**Yeast caspase 1 promotes the damage of green fluorescent protein positioned in the inner mitochondrial membrane during ROS burst**

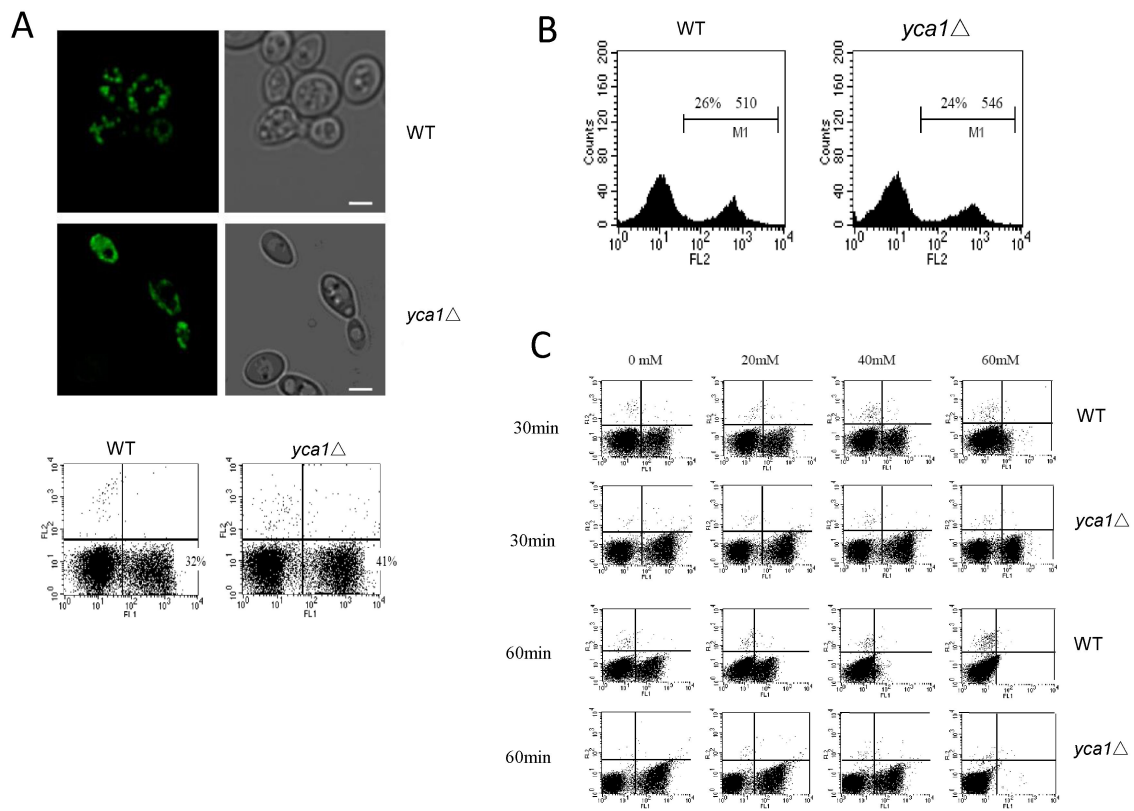

Fig. 4 A. Wild type and *yca1*Δ BY4741 cells with GFP-labeled mitochondria. Cells was observed with confocal microscope（Bar, 5μm.）,and analyzed with FACS (the PI fluorescence was detected by FL2 channel).

B. Comparison of superoxide anion induced by formic acid and detected with DHE in wild type and *yca1*Δ ATPase-GFP cells. The experiment was performed at least three times, with similar results.

C. Effect of formic acid induced ROS burst on the fluorescence intensity of ATPase-GFP in wild type and *yca1*Δ cells in stationary phase. The cells were treated for 30 min and 60 min separately, with formic of 0mM, 20mM, 40mM, 60mM, and PI staining (FL2 channel). The experiment was performed at least three times, with similar results.

We next investigated why the knockout of *YCA1* gene or the inhibition of Yca1 activity intensified the ROS burst. As Yca1 is a protease, we speculated that Yca1 accelerated protein degradation and increased the intracellular concentration of reductive small molecules to suppress ROS more effectively. In order to observe hydrolysis by caspsase, we fused GFP to an ATPase located in the inner mitochondrial membrane of the cell and used fluorescence as an indicator of GFP structural change.

GFP fluorescence in wild type and *yca1* mutant cells showed the shape of the mitochondria clearly (Fig 4A), indicating the specific binding of GFP to the mitochondrial ATPase.

With the same concentration (60 mM) of formic acid treatment, the superoxide anion levels (detected with DHE) are not obviously different between the ATPase-GFP-containing *yca1* cells compared to the ATPase-GFP-containing wild type control (26% and 24 % ) (Fig 4B). Thus, GFP-labeled wild type and *yca1* mutant cells have a similar ability to produce superoxide anion after formic acid treatment.

Figure 4C shows quickly decreasing GFP fluorescence intensity during ROS burst. We also found that the GFP fluorescence in wild type cells disappeared faster than in *yca1* mutant cells, indicating that GFP structure was damaged faster in the wild type cells. Therefore, Yca1 promoted rapid GFP destruction or degradation.

Because the GFP is bound and destruction appears to occur rapidly, the GFP is likely oxidized and degraded directly, rather than being degrading after misfolding or aggregation.

## 4. Discussion and Conclusions

Madeo etc. first identified metacaspase Yca1 in *S.cerevisiae* (Madeo, Herker et al. 2002), also called Mca1 (Vercammen, Declercq et al. 2007). Recent evidence suggested that yeast metacaspase is involved in some physiological processes in addition to cell apoptosis: (1) Over-expression of metacaspase in the fission yeast *Schizosaccharomyces pombe* can promote cell growth and enhance resistance of cells to hydrogen peroxide (Lim, Kim et al. 2007); (2) caspase inhibition improves cell mortality and leads to generation of ROS (May and Madge 2007); and perhaps most interesting, (3) the levels of oxidized proteins after $H_2O_2$ treatment are higher in a *YCA1* null strain than in a wild type control (Khan, Chock et al. 2005). Although this result can be explained if the *YCA1* null cell is able to endure more oxidative stress without the process of apoptosis, resulting in more oxidized proteins, this finding could also be considered as Yca1 playing a role in anti-oxidation.

There has been growing awareness of the physiological significance of Yca1 (Galluzzi, Joza et al. 2008), but a full understanding of its multiple roles in cell and determining the meaning of apoptosis to single-cell organism has remained elusive.

We studied formic acid induced yeast apoptosis to advance our understanding of yeast apoptosis and the physiological function of caspase. We found that formic acid induced a rapid ROS burst, which was even faster in cells lacking Yca1 (Du, Su et al. 2008). Cells have multiple mechanisms to eliminate ROS, such as using SOD enzymes and reductive substances such as glutathione to quench free radicals, but these reductive substances are easily consumed while generation of ROS can be quite robust (Macho, Hirsch et al. 1997; Buonocore, Perrone et al. 2010). Therefore, we propose that when the intracellular level of ROS reaches a certain level, and strong reduction components such as glutathione have been depleted, Yca1 is activated to directly hydrolyze or indirectly promote the hydrolysis of protein, especially surface-oxidized proteins. The amino acid residue specificity and cleavage point specificity of this enzyme will allow formation of a hydrophilic double carboxyl terminal. As a consequence, the structure of the oxidized protein may be destabilized and more susceptible to further oxidation. These caspase-hydrolyzed proteins may be then labeled by ubiquitin or disaggregated by HSP, and then completely hydrolyzed by the proteasome. The hydrolysis product, amino acid or polypeptide, which is more reductive than protein, can help to eliminate ROS faster. This activation of caspase may lead to other protective processes, such as cell shrinkage and chromatin condensation, which can also protect the cell from oxidative damage. This hypothesis that caspase accelerates degradation of oxidized protein may also help to explain some previously published findings. Both in human and yeast cells, Khan and coworkers observed that caspase deletions caused increased levels of intracellular oxidized protein after $H_2O_2$ treatments (Khan, Chock et al. 2005). More recently, Yca1 was found to accelerate the clearance of insoluble protein aggregates and to regulate the composition of the insoluble proteome (Lee, Brunette et al. 2010; Shrestha, Puente et al. 2013).Cells of *yca1* mutant cultured in an oxidative stress condition had lower respiration activities than wild type cells (Lefevre, Sliwa et al. 2012). Finally, recent reports show that elevating *YCA1* (*MCA1*) expression facilitates the removal of unfolded protein, counteracts accumulation of protein aggregates, and prolongs cellular life span in an Hsp104 disaggregase- and proteasome-dependent manner (Hill, Hao et al. 2014; Kampinga 2014).

We tested the hypothesis that Yca1 accelerates the hydrolysis of oxidized protein during ROS burst and found that formic acid induced an even faster ROS burst in cells lacking *YCA1* or inhibited for Yca1 enzyme activity. We also bound fluorescent protein (GFP) to the inner mitochondrial membrane and observed its structural change during ROS burst. GFP is very stable and resistant to hydrolysis (Roucou, Prescott et al. 2000), but GFP fluorescence disappears rapidly during ROS burst, indicating rapid changes to GFP structure. GFP fluorescence disappeared more slowly without Yca1 activity, indicating that Yca1 promotes the hydrolysis of GFP.

Oxidative stress is closely related to the maintenance of mitochondrial activity. If Yca1 reduces oxidative damage, it should also contribute to the maintenance of mitochondria activity. We found that mitochondrial activity of cells lacking the *YCA1* gene did not significantly change during log phase. However, in aging cells, the mitochondrial activity of *yca1* mutant cells decreased faster, indicating Yca1 helped to maintain mitochondrial activity. Because Yca1 can attenuate the burst of ROS, we speculate that the cell aging process and burst of the ROS are similar processes of oxidative damage and that Yca1, which slows the ROS burst, is also able to reduce mitochondrial damage during aging.

Yca1 may affect mitochondria activity. Mitochondria are a central part of aerobic metallization and also play an important role in cell apoptosis. In mice, caspase-2 gene deletion decreases the ability to remove oxidation proteins and has a significant impact on cellular senescence (Zhang, Padalecki et al. 2007). Knockout of the *YCA1* gene or inhibition of the corresponding enzyme prolongs the G1/S phase of cell cycle, slowing down cell growth in fermentation conditions (Lee, Puente et al. 2008).

During the aging process, the fact that mitochondrial activity of the *Yca1* mutant cell is lower than the wild type cells may provide another explanation for the finding that *YCA1* knockout reduces the sensitivity of cells to the stresses of hydrogen peroxide and acetic acid. *(Madeo, Herker et al. 2002; Guaragnella, Pereira et al. 2006)*. A lower mitochondrial activity of *yca1* mutant cells can slow metabolism and nutrition consumption may explain why these cells survive longer but are unable to grow as fast as wild type cells in a competitive condition (Herker, Jungwirth et al. 2004).

In addition to Yca1, we speculate that other caspases may accelerate the hydrolysis of oxidized proteins and decrease oxidative damage during ROS burst. The anti-oxidation function of caspase we suggest here is consistent with the mitochondrial origin of caspase (Kroemer 1997; Boyce, Degterev et al. 2004). Caspase, an enzyme that may be an adaptive result of aerobic metabolism, originated in aerobic bacteria, transferred to eukaryotic cells with mitochondria by symbiosis, and then evolved as a component of apoptosis. Our results suggest that the primary function of caspase may be closely linked to the response to oxidation stress in aerobic metabolism.

## Acknowledgments

Thanks for the financial support (2011B090400452) from Science and Technology Department of Guangdong Province in China.